\documentclass{webofc}
\usepackage[varg]{txfonts}   
\usepackage[english]{babel}
\usepackage{hyperref}
\usepackage{bm}
\usepackage{bbold}
  \def\pa{\partial}

\def\pda{\overleftrightarrow{\partial}}
\def\lQ{\Lambda_{\rm QCD}}
\begin{document}
\title{Doubly heavy baryons in Born-Oppenheimer EFT}

\author{\firstname{Jaume} \lastname{Tarr\'us Castell\`a}\inst{1,2}\fnsep\thanks{\email{jtarrus@iu.edu}}    
}

\institute{Department of Physics, Indiana University, Bloomington, Indiana 47401, USA
\and
Center for Exploration of Energy and Matter, Indiana University, Bloomington, Indiana 47408, USA}

\abstract{We report on the recent progress on the computation of the doubly heavy baryon spectrum in effective field theory. The effective field theory is built upon the heavy-quark mass and adiabatic expansions. The potentials can be expressed as NRQCD Wilson loops with operator insertions. These are nonperturbative objects and so far only the one corresponding to the static potential has been computed with lattice QCD. We review the proposal for a parametrization of the potentials based in an interpolation between the short- and long-distance regimes. The long-distance description is obtained with a newly proposed Effective String Theory which coincides with the previous ones for pure gluodynamics but it is extended to contain a fermion field. We show the doubly heavy baryon spectrum with hyperfine contributions obtained using these parametrizations for the hyperfine potentials.}
\maketitle
\section{Introduction}\label{intro}

In the past two decades, there have been continuous discoveries at various experimental facilities of hadrons that fall into the category of doubly heavy hadrons. In this category one can find the nearly two dozen exotic quarkonium states discovered so far, such as the double charm tetraquark, $T^+_{cc}$, recently discovered by the LHCb Collaboration \cite{LHCb:2021vvq}, as well as the somewhat older discovery of the pentaquark states~\cite{Aaij:2015tga}. Furthermore, traditional hadrons, in the quark model sense, also belong to this class, such as standard quarkonium and the main subject of this report, doubly heavy baryons, which have been first observed by the LHCb Collaboration~\cite{Aaij:2017ueg,Aaij:2018gfl} in the form of the $\Xi^{++}_{cc}$.

All doubly heavy hadrons share a set of characteristic scales. The ratios of some of these scales are small and can be used as expansion parameters for the description of the properties of these states. Two of these expansion can be applied to almost all of the cases of interest. First, the heavy-quark mass, $m_Q$, is larger than the characteristic scale of momentum transfer in hadrons, $\lQ$, which is the scale of nonperturbative dynamics in QCD. The second is that one can perform an adiabatic expansion between the heavy-quark dynamics and the one of the light degrees of freedom, that is the gluons an light quarks. The characteristic energy scale of the heavy-quark dynamics is their binding energy, which can be written as $m_Q v^2$, with $v\ll 1$ the heavy-quark relative velocity, while the light degrees of freedom dynamics are dominated by nonperturbative effects and therefore characterized by the $\lQ$ scale.

An Effective Field Theory (EFT) describing any doubly heavy hadron\footnote{Excluding states corresponding to bound states of pairs of singly heavy hadrons.} exploiting these two expansion has been proposed in Ref.~\cite{Soto:2020xpm}. Since this EFT reproduces the Born-Oppenheimer (BO) approximation at leading order it is often referred as BOEFT. The EFT was constructed in the single hadron sector up to the heavy-quark spin and angular momentum terms suppressed by $1/m_Q$. Expressions of the potentials as operator insertions in the Wilson loop were obtained by matching the EFT to nonrelativistic QCD (NRQCD)~\cite{Caswell:1985ui,Bodwin:1994jh,Manohar:1997qy}. The computation of the Wilson loop with operator insertions cannot be done using perturbative techniques and should be carried out (ideally) in lattice QCD or other nonperturbative approaches. 

BOEFT has been applied to doubly heavy baryons in Ref.~\cite{Soto:2020pfa}. In this case, the Wilson loop with light-quark operator insertions corresponding to the static potential, has been obtained in the lattice~\cite{Najjar:2009da,Najjarthesis} including several excited states. This lattice data was used in Ref.~\cite{Soto:2020pfa} to obtain the double charm and bottom baryon spectrum at leading order in BOEFT. In the case of the heavy-quark spin and angular-momentum dependent potentials there is no available lattice data. To sidestep this problem, in Ref.~\cite{Soto:2021cgk}, a phenomenological parametrization of these potentials was put forward. This parametrization consists of an interpolation between the descriptions of the potentials in the short- and long-distance regimes. In the short-distance regime, $r\ll \lQ^{-1}$, one can use the multipole expansion to find model-independent parametrizations of the potentials in terms of some nonperturbative constants. On the other hand, for the description of the potentials in the long-distance regime, $r\ll \lQ^{-1}$, an Effective String Theory (EST)~\cite{Nambu:1978bd}  has been developed in  Ref.~\cite{Soto:2021cgk}. This EST is an extension of the one which has been used to very successfully describe the standard and hybrid quarkonium potentials~\cite{PerezNadal:2008vm,Brambilla:2014eaa,Hwang:2018rju,Oncala:2017hop}.

\section{BOEFT for doubly heavy baryons}\label{sec:1}

Doubly heavy baryons are formed by two distinct components: a heavy quark pair and a light quark. At leading order in NRQCD the heavy quarks are static and the spectrum of the theory is given by the so-called static energies. These are characterized by a set of quantum numbers: the flavor of the light quark, the heavy quark pair relative distance $\bm{r}$, and the representation of $D_{\infty h}$\footnote{See Ref.~\cite{Berwein:2015vca} for a detailed discussion on the representations of $D_{\infty h}$.}. The lattice results in Refs.~\cite{Najjar:2009da,Najjarthesis} show that the lowest lying static energy corresponds to the representation $(1/2)_g$ followed by three very close states corresponding to the representations $(1/2)_u$, $(3/2)_u$ and $(1/2)_u'$. In the short-distance limit the symmetry group is enlarged from $D_{\infty h}$ to $O(3)$ and the states can be labeled by their spin ($\kappa$) and parity ($p$). Projecting the $\kappa^p$ states into the heavy quark axis one can obtain states in representations of $D_{\infty h}$. We show the correspondence in Table~\ref{repcor}. Here we will consider only doubly heavy baryon states associated to the $\kappa^p=(1/2)^{\pm}$ light quark states.

\begin{table}[ht!]
\centering
\begin{tabular}{||c|c||} \hline\hline
$O(3)$ & $D_{\infty h}$ \\\hline
$(1/2)^+$ & $(1/2)_g$ \\
$(3/2)^-$ & $(1/2)_u,\,(3/2)_u$ \\
$(1/2)^-$ & $(1/2)_u'$ \\\hline\hline
\end{tabular}
\caption{Correspondence between short-distance $O(3)$ representations and $D_{\infty h}$ representations.}
\label{repcor}
\end{table}

The Hamiltonian densities associated to the $\kappa^p=(1/2)^{\pm}$ light-quark states~\cite{Soto:2020pfa} have the following expansion up to $1/m_Q$
\begin{align}
h_{(1/2)^{\pm}}=\frac{\bm{p}^2}{m_Q}+\frac{\bm{P}^2}{4m_Q}+V_{(1/2)^{\pm}}^{(0)}(\bm{r})+\frac{1}{m_Q}V_{(1/2)^{\pm}}^{(1)}(\bm{r},\,\bm{p})\,.\label{hamden}
\end{align}
At leading order we have just the static potential
\begin{align}
V_{(1/2)^{\pm}}^{(0)}(\bm{r})=&V_{(1/2)^{\pm}}^{(0)}(r)\,.\label{lopot}
\end{align}
The heavy-quark spin and angular-momentum dependent operators appear at next-to-leading order and read as
\begin{align}
V_{(1/2)^{\pm}{\rm SD}}^{(1)}(\bm{r})=&V^{s1}_{(1/2)^{\pm}}(r)\bm{S}_{QQ}\cdot\bm{S}_{1/2}+V^{s2}_{(1/2)^{\pm}}(r)\bm{S}_{QQ}\cdot\left(\bm{{\cal T}}_{2}\cdot\bm{S}_{1/2}\right)+V^{l}_{(1/2)^{\pm}}(r)\left(\bm{L}_{QQ}\cdot\bm{S}_{1/2}\right)\,,\label{nlopot}
\end{align}
with ${\cal T}^{ij}_2=\hat{\bm{r}}^i\hat{\bm{r}}^j-\delta^{ij}/3$, $\bm{S}_{1/2}=\bm{\sigma}/2$ and $2\bm{S}_{QQ}=\bm{\sigma}_{QQ}=\bm{\sigma}_{Q_1}\mathbb{1}_{2\,Q_2}+\mathbb{1}_{2\,Q_1}\bm{\sigma}_{Q_2}$, where $\bm{\sigma}$ are the standard Pauli matrices and $\mathbb{1}_2$ is an identity matrix in the heavy-quark spin space for the heavy quark labeled in the subindex.

For the potentials in Eqs.~\eqref{lopot} and \eqref{nlopot} the matching expressions of Ref.~\cite{Soto:2020xpm} in terms of the Wilson loop with operator insertions reduce to
\begin{align}
V_{(1/2)^{\pm}}^{(0)}(\bm{r})&=\lim_{t\to\infty}\frac{i}{t}\log\left({\rm Tr}\left[\langle1\rangle^{(1/2)^{\pm}}_{\Box}\right]\right)\,,\label{lopotst}
\end{align}
and
\begin{align}
V^{s1}_{(1/2)^{\pm}}(r)&=-c_F\lim_{t\to\infty}\frac{4}{3t}\int^{t/2}_{-t/2} dt^{\prime}\frac{{\rm Tr}\left[\bm{S}_{1/2}\cdot\langle g\bm{B}(t^{\prime},\bm{x}_1)\rangle^{(1/2)^{\pm}}_{\Box}\right]}{{\rm Tr}\left[\langle1\rangle^{(1/2)^{\pm}}_{\Box} \right]}\,,\label{nlopots1}\\
V^{s2}_{(1/2)^{\pm}}(r)&=-c_F\lim_{t\to\infty}\frac{6}{t}\int^{t/2}_{-t/2} dt^{\prime}\frac{{\rm Tr}\left[\left(\bm{S}_{1/2}\cdot \bm{{\cal T}}_{2}\right)\cdot\langle g\bm{B}(t^{\prime},\bm{x}_1)\rangle^{(1/2)^{\pm}}_{\Box}\right]}{{\rm Tr}\left[\langle1\rangle^{(1/2)^{\pm}}_{\Box} \right]}\,,\label{nlopots2}\\
V^{l}_{(1/2)^{\pm}}=&-\lim_{t\to\infty}2\int^{1}_{0} ds\,s\frac{{\rm Tr}\left[\bm{S}_{1/2}\cdot\left(\frac{2}{3}\mathbb{1}_2 -\bm{{\cal T}}_{2}\right)\cdot\langle g\bm{B}(t/2,\bm{z}(s))\rangle^{(1/2)^{\pm}}_{\Box}\right]}{{\rm Tr}\left[\langle1\rangle^{(1/2)^{\pm}}_{\Box} \right]}\,,\label{nlopotl}
\end{align}
where $\bm{z}(s)=\bm{x}_1+s(\bm{R}-\bm{x}_1)$ and we use the following notation for the Wilson loop averages
\begin{align}
&\langle \dots\rangle^{(1/2)^{\pm}}_{\Box}=\langle {\cal Q}_{(1/2)^{\pm}}(t/2,\,\bm{R})\dots {\cal Q}_{(1/2)^{\pm}}^{\dagger}(-t/2,\,\bm{R})P\left\{e^{-ig\int_{{\cal C}_1+{\cal C}_2}dz^{\mu}A^{\mu}(z)}\right\}\rangle\,,
\end{align}
with ${\cal C}_1$ and ${\cal C}_2$ the upper and lower paths of a rectangular Wilson loop. Note that, unlike the quark-antiquark case, the flow is in the same direction for both paths. The interpolating operators are
\begin{align}
{\cal Q}^{\alpha}_{(1/2)^+}(t,\bm{x})&=\left[P_+q^{l}(t,\bm{x})\right]^\alpha\underline{T}^l \,,\\ 
{\cal Q}^{\alpha}_{(1/2)^-}(t,\bm{x})&=\left[P_+\gamma^5q^l(t,\bm{x})\right]^\alpha\underline{T}^l \,,
\end{align}
where $\alpha=-1/2,1/2$, and we have used the following $\bar{3}$ tensor invariants 
\begin{align}
&\underline{T}^l_{ij}  = \frac{1}{\sqrt{2}} \epsilon_{lij},\quad i,\,j,\,l=1,2,3\,.
\end{align}

\section{Short-distance regime}\label{sec:2}

\begin{figure}[ht!]
   \centerline{\includegraphics[width=.9\textwidth]{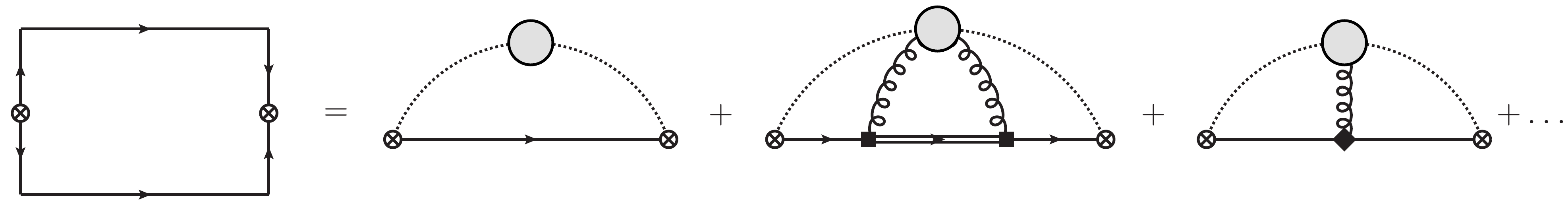}}   \centerline{\includegraphics[width=.99\textwidth]{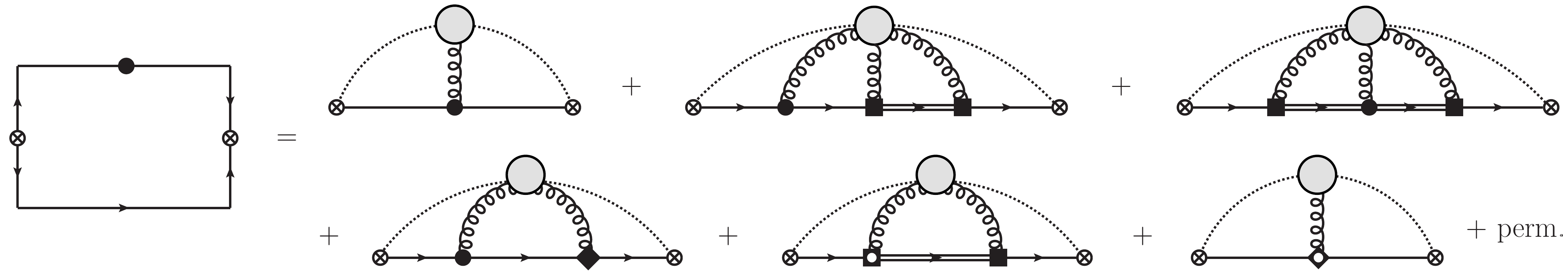}}
	\caption{Matching of the Wilson loop with operators insertions with the expansion in weakly-coupled pNRQCD up to next-to-leading order. The single lines represent the antitriplet fields, the double lines the sextet field, the dotted and the curly lines the light-quark and gluon fields respectively. The crossed circles indicate the insertion of a ${\cal Q}$ operator while the other vertices stand for various multipolar couplings of the heavy-quark pair with chromoelectric or chromomagnetic fields. The upper line corresponds to the static potential while the second and third correspond to the heavy-quark spin-dependent potentials.}
	\label{sp_matching}
 \end{figure}

In the short-distance regime $r\ll \lQ^{-1}$. Since the relative momentum of the heavy quarks scales as $m_Qv\sim r^{-1}$, in this regime it is a perturbative quantity. Therefore, one can integrate it out and build (weakly-coupled) potential NRQCD (pNRQCD)~\cite{Pineda:1997bj,Brambilla:1999xf,Brambilla:2005yk}. One can then integrate out the $\lQ$ modes and match pNRQCD to BOEFT. Using this two step matching procedure one finds multipole expanded expressions of the potentials. Other examples of this two-step matching can be found in Refs.~\cite{Brambilla:2018pyn,Brambilla:2019jfi} for the heavy-quark spin dependent potentials of quarkonium hybrids and in Refs.~\cite{Pineda:2019mhw,Castella:2021hul} for the hybrid and standard quarkonium transitions. 

The expansion of the static potential in Eq.~\eqref{lopotst} is given diagrammatically in Fig.~\ref{sp_matching} and corresponds to the following form
\begin{align}
V^{(0)}_{(1/2)^{\pm}}(r)&=-\frac{2}{3}\frac{\alpha_s}{r}+\overline{\Lambda}_{(1/2)^{\pm}}+\overline{\Lambda}^{(1)}_{(1/2)^{\pm}}r^2+\dots\,,\label{sdstp}
\end{align}
with $\overline{\Lambda}_{(1/2)^{\pm}}$ and $\overline{\Lambda}^{(1)}_{(1/2)^{\pm}}$ nonperturbative constants. The former can be obtained $D$ or $B$ meson masses from heavy quark-diquark duality~\cite{Savage:1990di,Hu:2005gf,Mehen:2017nrh,Mehen:2019cxn}. One can find analogous expansions for the hyperfine potentials:
\begin{align}
V^{s1}_{(1/2)^{\pm}}(r)&=c_F\left(\Delta^{(0)}_{(1/2)^{\pm}}+\Delta^{(1,0)}_{(1/2)^{\pm}}r^2+\dots\right)\,,\label{sdps1}\\
V^{s2}_{(1/2)^{\pm}}(r)&=c_F\Delta^{(1,2)}_{(1/2)^{\pm}}r^2+\dots\,,\label{sdps2}\\
V^{l}_{(1/2)^{\pm}}=&\frac{1}{2}\left[\Delta^{(0)}_{(1/2)^{\pm}}+\left(\Delta^{(1,0)}_{(1/2)^{\pm}}-\frac{1}{3}\Delta^{(1,2)}_{(1/2)^{\pm}}\right)r^2\right]+\dots\,,\label{sdpl}
\end{align}
with the $\Delta$'s corresponding to nonperturbative constants which expression as pNRQCD correlators can be found Ref.~\cite{Soto:2021cgk}. The constants $\Delta^{(0)}_{(1/2)^{\pm}}$ can be obtained using the heavy quark-diquark duality from the hyperfine splittings of heavy mesons.

\section{Long-distance regime}\label{sec:3}

At long distances a flux tube emerges from the heavy quarks joining at the position of the light quark. The dynamics of the flux tube can be described by an EST which coincides with the one for pure gluodynamics with the addition of a fermion field constrained to move on the string.

The action of the gluonic string is just proportional to the area of the string world sheet
\begin{align}
S_{\rm g}=-\sigma\int d^2 x\sqrt{|{\rm det}g_{ab}|}\,,\label{gst}
\end{align}
with $\sigma$ the string tension, and $g_{ab}$ the induced metric on the string. The action of a four-dimensional Dirac field constrained on a string is given by
\begin{align}
S_{\rm l.q}=\int d^2 x\sqrt{g}\bar{\psi}(x)\left(i\rho^a\pda_a-m_{\rm l.q.}\right)\psi(x)\quad,\quad \bar{\psi}\rho^a\pda_a\psi\equiv\left(\bar{\psi}(\rho^a\pa_a\psi)-(\pa_a\bar{\psi})\rho^a\psi\right)/2
\label{ngaplq}
\end{align}
with $\rho^a\equiv \gamma^{\mu}e^a_{\mu}$, and $e^{\alpha}_a\equiv\partial \xi^{\alpha}/\partial x^a$ the Zweibein. The antisymmetrization of the partial derivative is required by Hermiticity.

Expanding the action in Eq.~\eqref{gst} for small string fluctuations $\pa\xi\sim (r\Lambda_{\rm QCD})^{-1}\ll 1$ we arrive at
\begin{align}
S_{\rm g}=-\sigma \int dt dz\left(1-\frac{1}{2}\pa^a\xi^l\pa_a\xi^l+\dots\right)\,,\label{gstex}
\end{align}
and for the case of the fermionic action in Eq.~\eqref{ngaplq} we find
\begin{align}
S_{\rm l.q}&=\int  dt dz\left(\bar{\psi}(t,z)i\gamma^a\pda_a\psi(t,z)-m_{\rm l.q.}\bar{\psi}(t,z)\psi(t,z)-\pa^a\xi^l\bar{\psi}(t,z)i\gamma^l\pda_a\psi(t,z)+\dots\right)\,,\label{ngaplqexp}
\end{align}
with $l=1,2,$ and $a=0,3$.

The mapping of the chromomagnetic field into string fluctuations can be found in Ref.~\cite{PerezNadal:2008vm}. However, mappings into string fermion operators are now possible and in fact provide the leading order contribution to the potentials in Eqs.~\eqref{nlopots1}-\eqref{nlopotl}. This mapping is as follows:
\begin{align}
\bm{B}^l(t,\,z)&\mapsto\Lambda_f\bar{\psi}(t,\,z)\frac{\bm{\Sigma}^l}{2}\psi(t,\,z)\,,
\label{map:e3}\\
\bm{B}^3(t,\,z)&\mapsto\Lambda^{\prime}_f\bar{\psi}(t,\,z)\frac{\bm{\Sigma}^3}{2}\psi(t,\,z)\,,\label{map:e4}
\end{align}
with $\bm{\Sigma}={\rm diag}(\bm{\sigma},\,\bm{\sigma})$. To convert the two-dimensional spin operators in Eqs.~\eqref{nlopots1} and \eqref{nlopots2} into four-dimensional spin operators, we have used the prescription $\bm{S}_{1/2}\mapsto\,\bm{\Sigma}/2$.

We compute the potentials in Eqs.~\eqref{lopotst}-\eqref{nlopotl} as correlators in the EST using the mapping  defined by Eqs.~\eqref{map:e3}-\eqref{map:e4}. We find the following result for the static potential:
\begin{align}
V_{(1/2)^{\pm}}^{(0)}(\bm{r})&=\sigma r +E_1\,,
\end{align}
and analogously for the heavy-quark spin and angular-momentum dependent potentials one finds
\begin{align}
V^{s1}_{(1/2)^{\pm}}(r)&=\frac{c_F}{3r}\left(1\mp\frac{m_{\rm l.q.}}{E_1}\right)\left(\Lambda^{\prime}_f-2\Lambda_f\right)\,,\label{ldps1}\\
V^{s2}_{(1/2)^{\pm}}(r)&=\frac{c_F}{r}\left(1\mp\frac{m_{\rm l.q.}}{E_1}\right)\left(\Lambda^{\prime}_f+\Lambda_f\right)\,,\label{ldps2}\\
V^{l}_{(1/2)^{\pm}}(r)=&-\frac{1}{2r}\left(1\mp\frac{4}{\pi^2}\frac{m_{\rm l.q.}}{E_1}\right)\Lambda_f\,,\label{ldpl}
\end{align}
with $E_1=\sqrt{(\pi/r)^2+m^2_{\rm l.q.}}$.

\section{Spectrum with hyperfine contributions}\label{sec:4}

The leading order spectrum for doubly heavy baryons can be obtained by solving the Shr\"odinger equation with the static potential in Eq.~\eqref{lopotst} which has been computed on the lattice in Refs.~\cite{Najjar:2009da,Najjarthesis}. The results for the leading order spectrum can be found in Ref.~\cite{Soto:2020pfa}. Analogous lattice computations of the hyperfine potentials in Eqs.~\eqref{nlopots1}-\eqref{nlopotl} are not available. To address this problem a parametrization of these potentials has been proposed in Ref.~\cite{Soto:2021cgk}. The parametrization consists of an interpolation between the short- and  long-distance regime descriptions discussed in Sec.~\ref{sec:2} and Sec.~\ref{sec:3}, respectively. The proposed parametrizations of the hyperfine potentials are as follows:
\begin{align}
&V^{s1\,{\rm int}}_{(1/2)^\pm}=c_F\frac{\left(\Delta^{(0)}_{(1/2)^{\pm}}+\Delta^{(1,0)}_{(1/2)^{\pm}}r^2 \right)r^6_0+\frac{\left(\Lambda^{\prime}_f-2\Lambda_f\right)}{3}\left(1\mp\frac{m_{\rm l.q.}}{E_1}\right)r^5}{r^6+r^6_0}\,,\label{s1int}\\
&V^{s2\,{\rm int}}_{(1/2)^\pm}=c_F\frac{\Delta^{(1,2)}_{(1/2)^{\pm}}r^2 r^6_0+\left(\Lambda^{\prime}_f+\Lambda_f\right)\left(1\mp\frac{m_{\rm l.q.}}{E_1}\right)r^5}{r^6+r^6_0}\,,\\
&V^{l\,{\rm int}}_{(1/2)^\pm}=\frac{1}{2}\frac{\left[\Delta^{(0)}_{(1/2)^{\pm}}+\left(\Delta^{(1,0)}_{(1/2)^{\pm}}-\frac{1}{3}\Delta^{(1,2)}_{(1/2)^{\pm}}\right)r^2\right]r^6_0-\Lambda_f\left(1\mp\frac{4}{\pi^2}\frac{m_{\rm l.q.}}{E_1}\right)r^5}{r^6+r^6_0}\,.\label{lint}
\end{align}
Note that for $r_0=0$ we recover the long-distance potentials and for $r_0\to\infty$ we recover the short-distance potentials.

To obtain the unknown parameters in the interpolated potentials in Eqs.~\eqref{s1int}-\eqref{lint} a $\chi^2$ function was minimized. This function is constructed as the sum of the hyperfine splittings corresponding to the masses of the doubly heavy baryons from the lattice determinations of Refs~\cite{Briceno:2012wt,Namekawa:2013vu,Brown:2014ena,Alexandrou:2014sha,Bali:2015lka,Padmanath:2015jea,Alexandrou:2017xwd,Lewis:2008fu,Brown:2014ena,Mohanta:2019mxo}. The formulas for the hyperfine contributions to the masses of doubly heavy baryons for the states associated to the static energies $(1/2)_g$ and $(1/2)_u'$ can be found in Ref.~\cite{Soto:2020pfa}. 

\begin{figure}[ht!]
\centering
 \includegraphics[width=.5\textwidth]{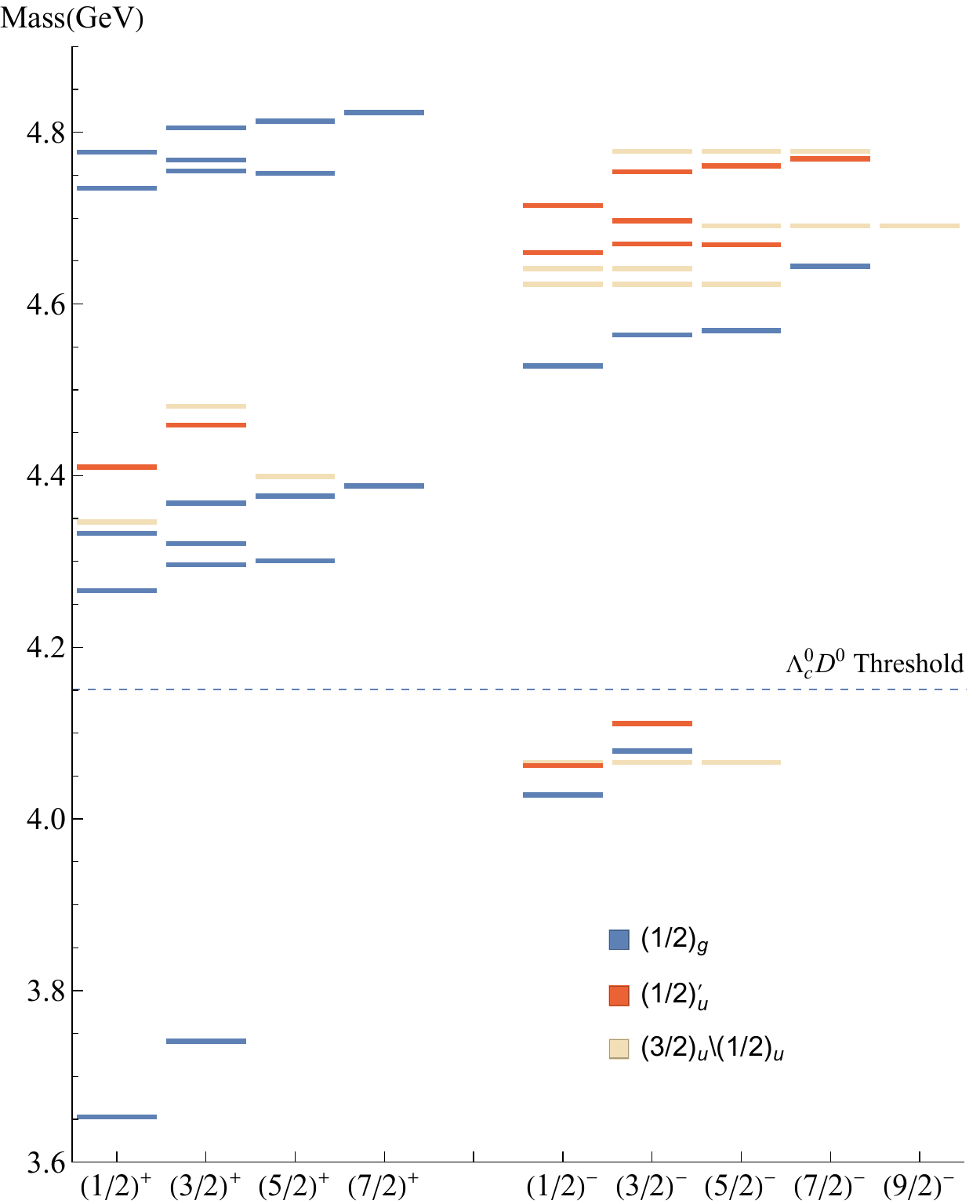}\includegraphics[width=.5\textwidth]{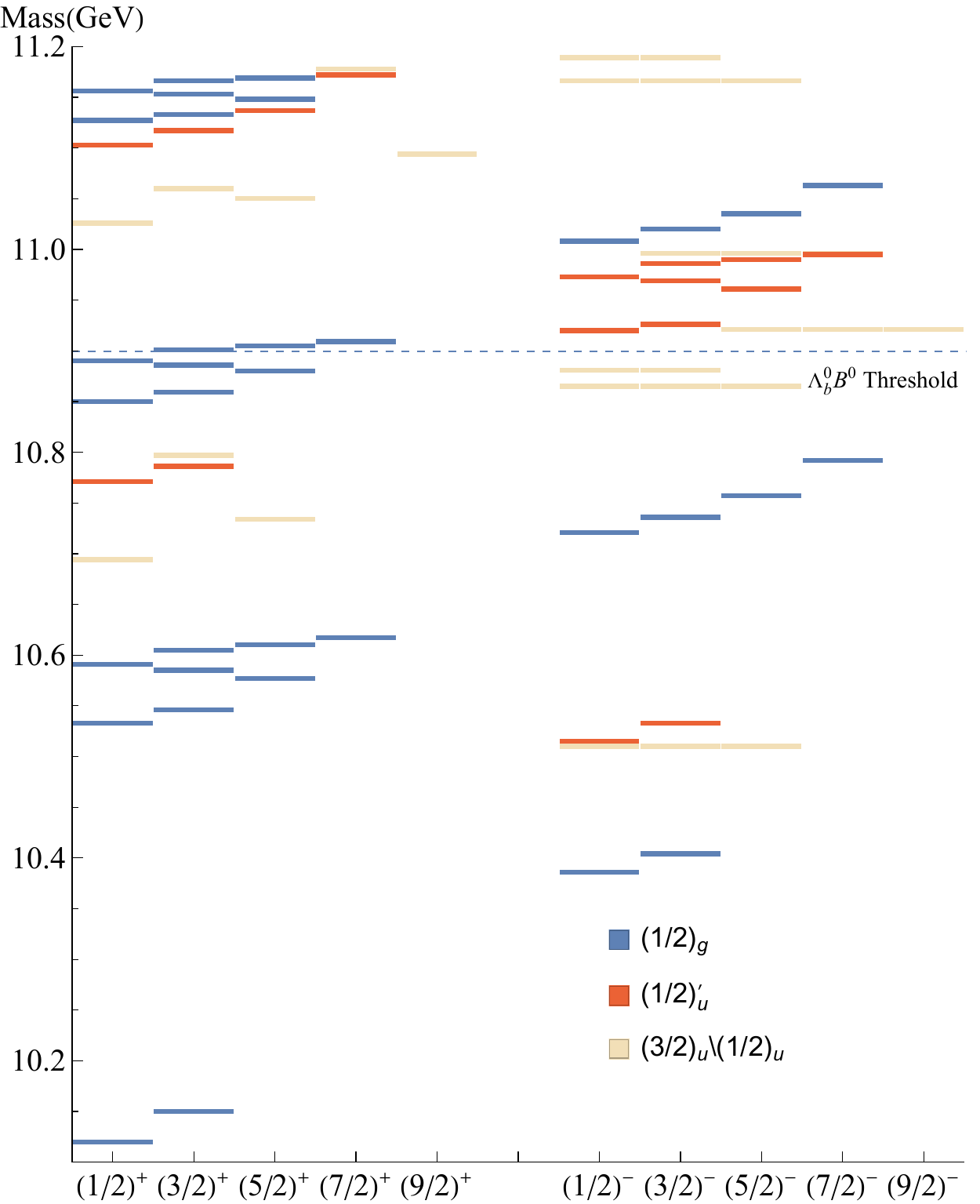}
	\caption{Spectrum of doubly heavy baryons in terms of $j^{\eta_P}$ states. Each line represents a state. The mixed $(3/2)_u\backslash(1/2)_u$states do not include hyperfine contributions}
	\label{ccplot}
 \end{figure}

Several fits where performed in Ref.~\cite{Soto:2021cgk}. The results of the fits show that with the current lattice data on the heavy baryon masses it is not possible to determine the values for $\Delta^{(1,0)}_{(1/2)^{\pm}}$ and $\Delta^{(1,2)}_{(1/2)^{\pm}}$. Therefore, the preferred fit consists of only the leading order short-distance piece with $\Delta^{(0)}_{(1/2)^{\pm}}$ fixed by heavy quark-diquark duality. The results of this fit can be found in table~\ref{fits2p} for various values of the interpolation parameter $r_0$.  In Fig.~\ref{ccplot} we show the spectrum of double charm and bottom baryons including the hyperfine contributions corresponding to the values of the parameters of $r_0=0.5$~fm in Table~\ref{fits2p}.

\begin{table}
\centering
\caption{Global fit of $\kappa^p=(1/2)^+$ $l=0,1,2,3$ multiplets hyperfine splittings for all the lattice data available for various values of $r_0$ with $\Delta^{(0)}_{(1/2)^{+}}=0.122$~{GeV}$^2$ from the $B$-meson splittings and $\Delta^{(1,0)}_{(1/2)^{+}}=\Delta^{(1,2)}_{(1/2)^{+}}=0$.}
\label{fits2p}
\begin{tabular}{cccc}\hline
$r_0~[{\rm fm}]$ & $\Lambda_f~[{\rm GeV}]$ & $\Lambda_f'~[{\rm GeV}]$ & $\chi^2_{\rm d.o.f}$\\\hline
$0.1$    & $-0.355(10)$ & $-0.265(19)$  & $0.66$ \\
$0.2$    & $-0.368(13)$ & $-0.264(25)$  & $0.72$ \\
$0.3$    & $-0.348(19)$ & $-0.270(33)$  & $0.69$ \\
$0.4$    & $-0.266(27)$ & $-0.286(44)$  & $0.60$ \\
$0.5$    & $-0.085(41)$ & $-0.314(61)$  & $0.58$ \\
$0.6$    & $ 0.224(75)$ & $-0.353(102)$ & $0.83$ \\  \hline
\end{tabular}
\end{table}

\section{Conclusions}

We have reported on the results for the doubly heavy baryon spectrum including hyperfine contributions. The hyperfine potentials, which can be written in terms of the Wilson loop with operator insertions, are nonperturbative quantities that should be, ideally, computed in lattice QCD. However, these computations are often not available. To sidestep this issue, in Ref.~\cite{Soto:2021cgk}, a parametrization of the potentials has been proposed. This consists of an interpolation between the short- and long-distance regimes. In the short-distance region the potentials can be computed in terms some unknown constants using weakly-coupled pNRQCD. Similarly, in the long-distance regime an Effective String Theory (EST) can be used. In Sec.~\ref{sec:3} we reviewed the EST with a fermion field constrained into the string which is appropriate to compute the potentials for doubly heavy baryons. This procedure to obtain reliable parametrizations of the potentials can be of significant utility in future studies of doubly heavy hadrons.

\section*{Acknowledgements}

J.T.C. acknowledges financial support by National Science Foundation (PHY-2013184). 

\bibliography{biblio}

\end{document}